\RequirePackage{fix-cm}
\documentclass[smallextended]{svjour3}       
\smartqed  
\usepackage{booktabs}
\usepackage{cite}
\usepackage{graphicx}
\usepackage{xcolor}
\usepackage{amsmath}
\usepackage{hyperref}
\usepackage{etoolbox}
\newcommand*{\affaddr}[1]{#1} 
\newcommand*{\affmark}[1][*]{\textsuperscript{#1}}
\begin{document}

\title{Two-Line Element Estimation Using Machine Learning
}


\author{Rasit Abay\protect\affmark[1] \and
        Sudantha Balage\affmark[1] \and
        Melrose Brown\affmark[1]  \and
        Russell Boyce\affmark[1]
}
\authorrunning{Rasit Abay \and Sudantha Balage\and Melrose Brown\and Russell Boyce}

\institute{   Rasit Abay \\
              \email{r.abay@student.unsw.edu.au}           
\\
\\
              \affaddr{\affmark[1]University of New South Wales, Canberra, Australian Capital Territory 2600, Australia}\\
}

\date{Received: date / Accepted: date}

\maketitle

\begin{abstract}
Two-line elements are widely used for space operations to predict the orbit with a moderate accuracy for 2-3 days. Local optimization methods, such as the nonlinear least squares method with differential corrections, can estimate a TLE as long as there exists an initial estimate that provides the desired precision. Global optimization methods to estimate TLEs are computationally intensive, and estimating a large number of them is prohibitive. In this paper, the feasibility of estimating TLEs using machine learning methods is investigated. First, a Monte-Carlo approach to estimate a TLE, when there are no initial estimates that provide the desired precision, is introduced. The proposed Monte-Carlo method is shown to estimate TLEs with root mean square errors below 1 km for space objects with varying area-to-mass ratios and orbital characteristics. Second, gradient boosting decision trees and fully-connected neural networks are trained to map orbital evolution of space objects to the associated TLEs using 8 million publicly available TLEs from the US space catalog. The desired precision in the mapping to estimate a TLE is achieved for one of the three test cases, which is a low area-to-mass ratio space object.
\end{abstract}

\section{Introduction}

Two-line element sets (TLEs) and Simplified General Perturbations \#4 (SGP4) are widely used for space operations to predict trajectories of space objects with moderate accuracy (tens of kilometers) for 2-3 days. The trajectories produced by SGP4 are defined in True Equator Mean Equinox (TEME) reference frame~\cite{DV2013}. The orbits of tracked Resident Space Objects around Earth are specified and updated in the US space catalog as TLEs. TLEs consist of mean elements that can only be used with SGP4 to propagate orbits with moderate accuracy. SGP4 provides the reference trajectory that is updated by fitting the actual trajectory, and TLEs are the updated parameters that approximate it. SGP4 includes zonal terms up to J5 of the gravitational potential of the Earth, neutral atmosphere with exponential decay and partially modeled third-body mass interactions~\cite{DV2013}. Bstar ($B^{*}$) is an element of a TLE that determines the effect of air drag on the trajectories of the space objects. Eq.~\ref{eq:BstarEq} shows the relationship between drag coefficient ($C_{D}$), area-to-mass ratio ($\frac{A}{m}$), atmospheric density ($\rho$) and $B^{*}$~\cite{DV2013}. However, it should be noted that the $B^{*}$ parameter in the TLE is highly sensitive to any perturbations, such as solar radiation pressure, air drag, third-body mass gravitational interactions, maneuvers and mismodeling of the Earth's gravitational potential. Therefore, $B^{*}$ is always adjusted during the estimation scheme of TLEs, and it can even be negative (7.9\% of the test data used in this work have negative $B^{*}$)~\cite{DV2013}.

\begin{equation}
\label{eq:BstarEq}
B^{*} = \frac{1}{2}\frac{C_{D}A}{m}\rho
\end{equation}

\noindent Mean motion ($n$) is one of the 6 independent mean elements, which include the inclination, right ascension of the ascending node, argument of perigee, mean anomaly and eccentricity,  that is required for computations of TLEs~\cite{DV2013}. Eq.~\ref{eq:EqMeanMotion} shows the relationship between the semi-major axis ($a$), the geocentric gravitational constant ($\mu$) and the mean motion of a space object. Mean motion has 8 decimal places in TLEs.

\begin{equation}
\label{eq:EqMeanMotion}
n = \sqrt{\frac{\mu}{a^{3}}}
  \end{equation}
  
  \noindent Inclination ($i$) is another independent mean element of a TLE that defines the orientation of the orbital plane with respect to the Earth. Inclination has 4 decimal places in a TLE, and its unit is degrees. The right ascension of the ascending node ($\Omega$) is a mean element of a TLE that defines the orientation of an orbit with respect to the $\hat{z}$ axis, which is parallel to the rotation axis of the Earth. It changes primarily due to the $J2$ perturbation that is caused by the oblateness of the Earth, and this change is a function of inclination. The right ascension of the ascending node has 4 decimal places in the TLE, and it is in degrees. Argument of perigee ($\omega$) is a mean element of a TLE that defines the orientation of the orbital plane with respect to the Earth around the axis parallel to the angular momentum vector of the orbit. The advance of the argument of perigee is the slowest frequency of the orbital evolution for objects orbiting the Earth while the change in the mean anomaly (Keplerian orbital frequency) is the fastest and the nodal precession is the intermediate frequency. Mean anomaly ($m$) is a mean element of a TLE that defines the position of a space object in its orbit. In addition, it is the fastest changing parameter. Eccentricity ($e$) is a mean element of a TLE that defines the shape of the orbit while semi-major axis ($a$) defines the size of the orbit. Eccentricity has 7 decimal places. The slow rate of change in the $a$ and $e$ of the low-Earth orbits is primarily due to air drag, which is a non-gravitational perturbation, and this is related to the orbital decay of the resident space objects in LEO.
  
  There are a few studies that investigate different approaches to estimate TLEs. As stated above, TLEs include the mean states estimated by fitting observations to the dynamics provided by SGP4, and they can only be used with SGP4~\cite{VC2006}. Keplerian orbital elements are used as initial estimates of TLEs by the differential corrections and nonlinear least squares methods~\cite{VC2008,JGMK1996,LBS2002}. Kalman filter is investigated to estimate TLEs using onboard Global Positioning System (GPS) data~\cite{MG2000}. However, the above methods, which search for the local minimum of the objective function, which is the sum of the squares of the position and velocity errors, depend on the availability of a reasonable initial estimate of a TLE. Although methods, such as genetic algorithms~\cite{GOH2018} and invasive weed optimization~\cite{BLD2015}, that do not require an initial estimate of the TLEs have been investigated, they are reported to be computationally intensive as they search for the global optimum. To summarize, the two main approaches in the literature to estimate a TLE in literature use either computationally expensive global search or local search which depends on having an initial estimate. This represents a significant shortcoming in the current state-of-the-art of orbit determination using TLEs. The present work addresses this shortcoming.
  
  This work utilizes machine learning methods, namely the gradient boosting trees and fully-connected neural networks, to approximate the inverse mapping of publicly available SGP4 algorithm~\cite{VC2006} for LEO objects by learning to map the orbital evolution to the associated TLE. The capability of approximating such mapping using machine learning will enable to represent time series orbital data in latent space that can be used with orbit propagators. The gradient boosting trees are machine learning models that are suitable for determining non-linear and sharp decision boundaries. The boosting is achieved by adding weak learners, such as decision trees, to determine the complex decision boundaries. The first successful application of such an idea is adaptive boosting (AdaBoost)~\cite{FS1995}. AdaBoost generates a sequence of classifiers, and each classifier puts more weight on the samples that are not classified accurately in the previous iteration. Such an approach enables each classifier to capture different features in the input data, and more complex decision boundaries can be defined in the end~\cite{KJ2013}. The boosting idea is further improved by defining the framework as a numerical optimization in the function space~\cite{Fr2001}. The objective function is minimized by adding weak learners using the steepest-descent method. 
  
  Gradient boosting of regression trees are shown to be efficient both for regression and classification problems~\cite{Fr2001}. The constraints imposed on trees and the weight updates, and the subsampling of data at each iteration help to reduce the overfitting~\cite{Fr2002}.
Moreover, L1 and L2 regularizations of the leaf weights in addition to the structure of the trees yield better generalization capability  because they force the gradient boosting decision trees to be less complex~\cite{CG2016}. 

Artificial neural networks are machine learning models that are suitable for determining smooth nonlinear decision boundaries. According to the universal approximation theory~\cite{Cy1989, HSW1989}, any Borel measurable function can be approximated by a feedforward neural network as long as it includes enough hidden units~\cite{GYC2016}. In addition, a neural network may also approximate any discrete function irrespective of its dimensions given that dimension is finite. Although the universal approximation theory states that any closed and bounded continuous function can be approximated by multilayer perceptrons, the training scheme to approximate that function might fail due to overfitting or failure of the optimization algorithm~\cite{GYC2016}.

\begin{figure}[!htbp]
\begin{center}
\includegraphics*[width=5cm]{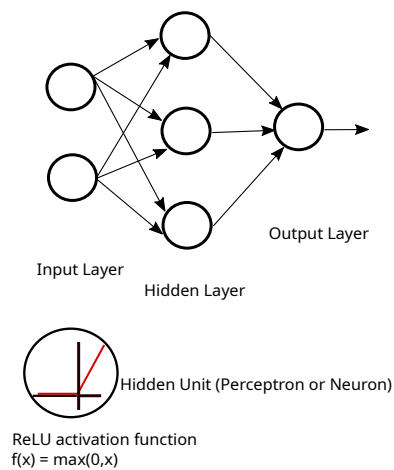}
\end{center}
\caption{Model of a simple neural network}
\label{figure4}
\end{figure}

Different techniques have been developed to solve the above issues regarding the neural network training, such as deeper networks, regularization and gradient descent with momentum. Figure~\ref{figure4} presents a simple neural network with an activation function of the rectified linear unit in the hidden layer. In this work, it is assumed that the publicly available SGP4 algorithm~\cite{VC2006} is a bounded and closed function, and the supervised machine learning models are investigated to approximate it. The contributions of the present study are : 1) a Monte-Carlo approach to estimate a TLE with the desired precision without an initial estimate as is required by the differential correction methods and 2) supervised machine learning methods which can map an orbital evolution to a TLE which can be used as an initial estimate with desired precision. The trained machine learning models can also be used to improve the computational efficiency of the proposed Monte-Carlo approach to estimate a TLE.

The paper is structured as follows: In section 2, a Monte-Carlo approach and supervised learning models to estimate TLEs are explained. In section 3, an investigation of the performances of the proposed methods is presented. In section 4, the discussion regarding the feasibility of using machine learning methods to approximate the inverse mapping of publicly available SGP4 algorithm~\cite{VC2006} for LEO objects is provided. In section 5, the conclusions and future work based on the results are discussed.

\section{Methodology}

\subsection{Monte-Carlo approach}

\begin{figure}[!htbp]
\begin{center}
\includegraphics*[width=13cm]{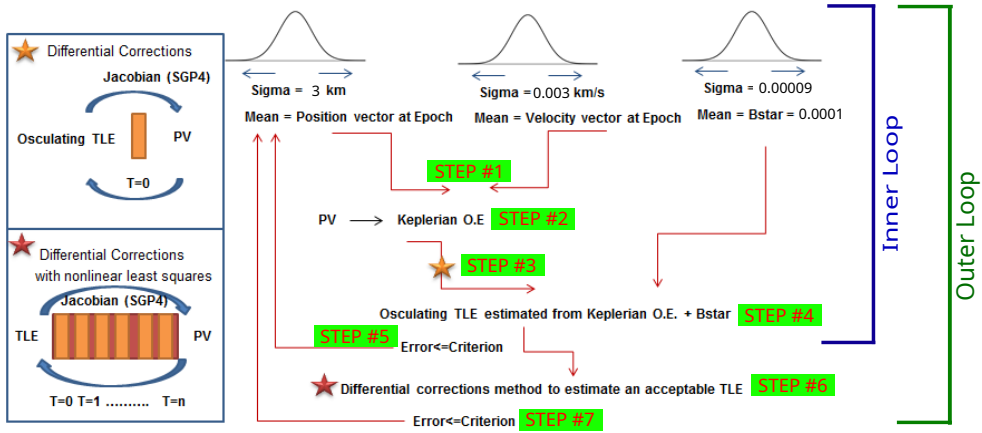}
\end{center}
\caption{Monte-Carlo approach to estimate a TLE}
\label{figure1}
\end{figure}

In this section, a Monte-Carlo approach that searches for the global minimum of the sum of the squares of the position and velocity errors to estimate a TLE without any initial data is presented. However, the proposed method is computationally intensive. Therefore, this work also investigates the feasibility of using machine learning to predict reasonable initial estimates of TLEs that can be utilized by the proposed Monte-Carlo approach. Figure~\ref{figure1} outlines the proposed method that extends the differential corrections method using the Monte-Carlo technique to estimate a TLE that satisfies a given criterion. The following procedure describes how to obtain such TLEs.

First, a position and a velocity vector are sampled from normal distributions which have the given position and velocity vector components (user-provided state for the epoch of the desired TLE) as mean values, and 3 km for the position and 0.003 km/s for the velocity as the associated standard deviations (STEP \#1 in Figure~\ref{figure1}). Then, the sampled position and velocity vectors are mapped to  the corresponding Keplerian elements (STEP \#2 in Figure~\ref{figure1}). Next, the mapped Keplerian elements are fed into SGP4 (Keplerian elements are assumed to be the initial estimate of the TLE with desired precision, and $B^{*}$ is taken as 0 at this step) to propagate to the epoch ($t=0$) to obtain new position and velocity vectors. The differential corrections to the Keplerian elements are computed by mapping the residual vector ($\delta rv_{epoch}$), which is the difference between the above obtained state vectors and the user-provided states, to the corrections ($\delta tle_{epoch}$) iteratively by using Eq.~\ref{eq:diffCorrection}   (STEP \#3 in Figure~\ref{figure1}). Once the process is converged (the magnitude of the position error vector is $10^{-8}$), an \textit{osculating} TLE is obtained, which provides the tangential orbit at the given point only ($t=0$). Next, the $B^{*}$ value ($\mu=0.0001$ and $\sigma=0.00009$) is sampled from the normal distribution computed from 12 years of all LEO TLEs in the official US space catalog (STEP \#4 in Figure~\ref{figure1}). The osculating TLE, which includes six independent elements that define the orbit, and the sampled $B^{*}$ are assumed to be the TLE with desired precision. The above computed TLE is used to propagate an orbit which  is compared to the orbit generated by high-precision numerical orbit propagator, which utilizes the libraries of Orekit (version 9.1) low level space dynamics library~\cite{MVP2010},  and the root mean square error is computed. The process starts from the beginning if the root mean square error (RMSE) does not satisfy the criterion, which is an RMSE of 30 km and smaller for this work to exclude the cases which may have secular error growth (STEP \#5 and the end of inner loop in Figure~\ref{figure1}). If the criterion at the end of the inner loop is satisfied, the iterative differential corrections and nonlinear least squares method is used to search for the TLE that corresponds to the local minimum by using Eq.~\ref{eq:diffCorrectionNonlinear} (STEP \#6 in Figure~\ref{figure1}). The sample that satisfies the error criterion (10 km and smaller in present study) is retained, and the process is terminated (STEP \#7 and the end of outer loop in Figure~\ref{figure1}). 

The differential corrections scheme iteratively improves the sampled osculating TLE to match the position vector at the epoch time. The stopping criterion is selected as $10^{-8}$ and smaller for the magnitude of position error in kilometers. Eq.~\ref{eq:diffCorrection} shows how a Jacobian matrix, which is computed by the finite-difference method, relates the change in the TLE to the change in the position and velocity vectors at the epoch.

\begin{equation}
\label{eq:diffCorrection}
\begin{split}
\delta tle_{epoch} &= \left(\frac{\partial f_{sgp4}}{\partial tle_{epoch}}\right)^{-1} \delta rv_{epoch}\\
&= (A_{t_{0}})^{-1} \delta b_{t_{0}}
\end{split}
\end{equation}

\noindent where $tle_{epoch}$ is the TLE parameters at the epoch, $f_{sgp4}$ is SGP4, $rv_{epoch}$ are the propagated position and velocity vectors at the epoch, $A_{t_{0}}$ is the Jacobian, and $\delta b_{t_{0}}$ is the residual at the epoch. Eq.~\ref{eq:diffCorrectionNonlinear} shows the normal equation that relates all of the changes in the position and velocity vectors associated with the observational data available to the change in the TLE at epoch time~\cite{VC2008}. In this work, all observational data are assumed to have the same weight, therefore, $W$ is assumed to be the identity matrix. Eq.~\ref{eq:diffCorrection} is used for differential corrections in the inner loop and Eq.~\ref{eq:diffCorrectionNonlinear} is used for differential corrections with nonlinear least squares method in the outer loop (Figure~\ref{figure1}).

\begin{equation}
\label{eq:diffCorrectionNonlinear}
\begin{split}
\delta tle_{epoch} &= (A^{T}WA)^{-1}A^{T}Wb\\
&= \left(\sum_{i}^{N} A_{(t_{i})}^{T}A_{(t_{i})}\right)^{-1} \sum_{i}^{N} A_{(t_{i})}^{T}b_{(t_{i})}
\end{split}
\end{equation}

\noindent where $tle_{epoch}$ is the TLE parameters at the epoch, $f_{sgp4}$ is SGP4, $rv_{epoch}$ are the propagated position and velocity vectors at any time, $A_{t_{i}}$ is the Jacobian, and $\delta b_{t_{i}}$ is the residual at any time.

\subsection{Supervised Machine Learning Models}

\subsubsection{Feature selection}

In the present work, Equinoctial orbital elements and a parameter that we shall call \textit{instance} (Table~\ref{tableInputData}) that defines the sequence of the data with fixed time interval are selected as feature vectors (input data for the machine learning models). The above orbital elements are non-singular, and the magnitude of their values are bounded, thus well-suited for the training process. The Keplerian mean motion, in radians per hour, replaces semi-major axis of the Equinoctial elements. The feature vectors are generated from 8,206,374 TLEs which are all LEO objects that have semi-major axes of 8,378 km and smaller in the space catalog. TLEs are restricted to the LEO orbital regime to ensure that all space objects are subject to a significant orbital perturbation due to atmospheric drag.

\begin{figure}[!htbp]
\begin{center}
\includegraphics*[width=10cm]{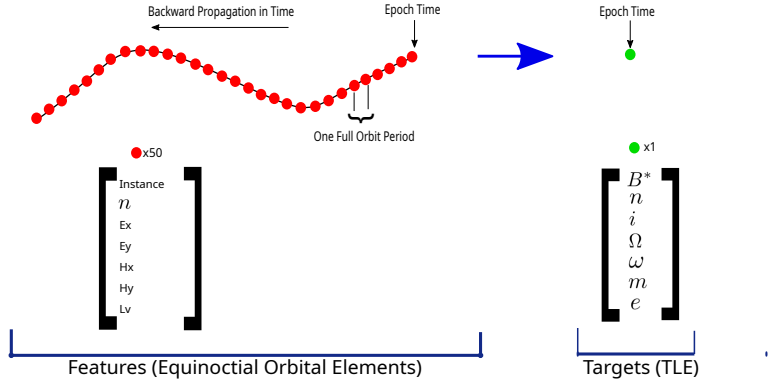}
\end{center}
\caption{Feature vectors}
\label{figure2}
\end{figure}

\begin{table}[h]
\center
\caption{Feature vectors}
\begin{tabular}{ll}
\hline
Features&Equations/Values\\
\hline
Instance & [2.452, 2.402, ..., 0.049, 0.0]\\
Mean motion (rad/h) & $n = \sqrt{\frac{\mu}{a^{3}}}$ \\
Ex & $E_{x} = ecos(\omega+\Omega)$ \\
Ey & $E_{y} = esin(\omega+\Omega)$  \\
Hx & $H_{x} = tan(\frac{i}{2})cos(\Omega)$  \\
Hy & $H_{y} = tan(\frac{i}{2})sin(\Omega)$  \\
Lv & $Lv = \nu + \omega + \Omega$  \\  
\hline
\end{tabular}
\label{tableInputData}
\end{table}

Figure~\ref{figure2} outlines the feature and target vectors. The dimensions of the feature and target vectors are 7x50 and 7x1 respectively. The target vectors are published TLEs in the official space catalog\footnote{Available at http://www.space-track.com.}. TLEs from the official space catalog are used for convenience because mappings for orbital regimes that are not populated by resident space objects are not desired. Therefore, TLEs are just used as initial conditions for SGP4 to generate associated time series orbital data. The feature vectors are generated by propagating the TLEs backward in time using SGP4 for 50 orbital periods and mapping each point to the associated Equinoctial orbital elements. Since TLEs are mostly computed from the history of orbital evolutions of the space objects which are obtained from past observations, the training and development test data include orbits that are propagated backward in time. The equinoctial orbital elements are selected as feature vectors because the Keplerian orbital elements have singularities and osculating TLEs are computationally expensive to obtain.

\subsubsection{Boosting trees}

An optimized distributed gradient boosting library called XGBoost (version 0.7) is used for this work\footnote{Available at https://github.com/dmlc/xgboost.}. A gradient boosting tree model is trained for each element of a TLE, namely Bstar ($B^{*}$), eccentricity ($e$), inclination ($i$), mean anomaly ($m$), mean motion ($n$), right ascension of the ascending node ($\Omega$), and the argument of perigee ($\omega$). The construction of individual gradient boosting trees for each TLE parameter is chosen because they cannot map to a multidimensional output for regression problems due to the architecture of the decision trees, whereas neural networks can. However, the above approach is expected to yield better function approximation because the complexity of the model is utilized to obtain one parameter only. Another disadvantage of XGboost is that it can not be efficiently used for incremental learning. All available data should be presented for constructing each tree. Therefore, the data should fit the memory to train the models. Due to the memory restriction (a workstation with 128 GB RAM and 64 AMD Opteron\textsuperscript{TM} 6376 processors are used), the training data include TLEs from 01 January 2017 to 21 April 2018, while test data include TLEs from 02 January 2016 to 01 April 2016 to select the best performing machine learning models. The models are trained with 8,206,374 TLEs and tested with 1,278,900 TLEs. A subset of the test data (69,632 TLEs from 27 March 2016 to 01 April 2016) is also used to determine the accuracy of the TLEs predicted by the selected best machine learning models. The test data are chosen from another time period as stated above to enable the trained machine learning models to generalize the predictions. Table~\ref{tableXgboostHyperparameters} shows the gradient boosting tree hyperparameters that are optimized for training XGBoost models. The learning rate controls the step size for the gradient-based optimizer. The number of estimators controls the number of trees. The maximum depth limits the depth of each tree which is used to increase the complexity of the model. The minimum child weight is required to avoid overfitting of the training data. A node is split as long as the resultant split leads to a positive reduction in the loss function (mean square error in this work). Gamma can control the split of nodes. The rest of the parameters in Table~\ref{tableXgboostHyperparameters} are used to avoid overfitting by forcing the model to be less complex. The hyperparameters of the gradient boosting trees are tuned empirically by considering the bias-variance tradeoff to ensure the trained models have generalization capability (Table~\ref{tableXgboostBiassVariance}). The grid search methods are not preferred because the models are trained with large amount of data.

\begin{table}[!htbp]
\center
\caption{Hyperparameters of XGBoost parameters}
\begin{tabular}{llllllll}
\hline
Parameters&$B^{*}$&$i$&$\Omega$&$e$&$\omega$&$m$&$n$\\
\hline
Learning rate & 0.1  & 0.3 & 0.3 & 0.3 & 0.2 & 0.2 & 0.1\\
No. of estimators & 100 & 80 & 70 & 70 & 80 & 80 & 100\\
Maximum depth & 20 & 20 & 20 & 20 & 25 & 25 & 25\\
Min. child weight & 7 & 9 & 11 & 13 & 27 & 27 & 1\\
Gamma & - & - & - & - & 0.7 & 1.0 & -\\
Subsample & 0.7 & 0.7 & 0.6 & 0.7 & 0.1 & 0.2 & 1.0\\
Col. sample by tree & 0.7 & 0.8 & 0.7 & 0.7 & 0.2 & 0.3 & 1.0\\
Col. sample by level & 0.7 & 0.8 & 0.7 & 0.7 & 0.2 & 0.3 & 1.0\\
Alpha & 0.2 & 0.2 & 0.2 & 0.1 & 0.9 & 0.7 & 0.0\\
\hline
\end{tabular}
\label{tableXgboostHyperparameters}
\end{table}

\begin{table}[!htbp]
\center
\caption{Bias-Variance tradeoff for XGBoost}
\begin{tabular}{lll}
\hline
Models&Training Squared Mean Error&Test Squared Mean Error\\
\hline
$B^{*}$ & 3.9e-5  & 6.3e-5\\
$i$ & 7.1e-6 & 2.1e-5\\
$\Omega$ & 3.5e-4 & 2.9e-3 \\
$e$ & 1.6e-6 & 1.3e-5 \\
$\omega$ & 0.21 & 0.32\\
$m$ & 0.15 & 0.26\\
$n$ & 9.8e-9  & 2.2e-8\\
\hline
\end{tabular}
\label{tableXgboostBiassVariance}
\end{table}

\begin{figure}[!htbp]
\begin{center}
\includegraphics*[width=4cm]{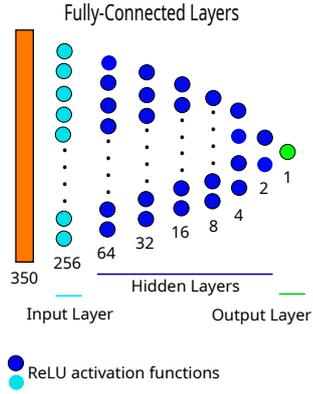}
\end{center}
\caption{Fully-connected neural networks}
\label{figure9}
\end{figure}

\begin{table}[h]
\center
\caption{Parameters to normalize and standardize the input data for the neural networks}
\begin{tabular}{lllll}
\hline
Features&Min&Max&$\mu$&$\sigma$\\
\hline
Instance & 0.0 & 2.45 & 1.225 & 0.721543\\
Mean motion (rad/h) & 1.532504 & 4.479097 & 3.662984 & 0.221211\\
Ex & -0.489632 & 0.430255 & -0.000028 & 0.015722\\
Ey & -0.327102 & 0.434307 & 0.000077 & 0.015767\\
Hx & -3.138717 & 3.138880 & 0.001386 & 0.693990\\
Hy & -3.138954 & 3.138859 & 0.009009 & 0.7005792\\
Lv & -3.141592 & 3.141592 & 0.01232161 & 1.809887\\  
\hline
\end{tabular}
\label{tableMinMaxScaling}
\end{table}

\begin{table}[!htbp]
\center
\caption{Fully-connected neural network hyperparameters}
\begin{tabular}{llllllll}
\hline
Parameters&$B^{*}$&$i$&$\Omega$&$e$&$\omega$&$m$&$n$\\
\hline
Learning rate & 9e-5  & 9e-5 & 9e-5 & 9e-5 & 9e-5 & 9e-5 & 1e-5\\
Optimizer & Adam & Adam & Adam & Adam & Adam & Adam & Adam\\
$\beta_{1}$ & 9e-4 & 9e-4 & 9e-4 & 9e-4 & 9e-4 & 9e-4 & 9e-4\\
$\beta_{2}$ & 999e-6 & 999e-6 & 999e-6 & 999e-6 & 999e-6 & 999e-6 & 999e-6\\
Decay & 1e-07 & 1e-07 & 1e-07 & 1e-07 & 1e-07 & 1e-07 & 1e-07\\
Epoch number & 50 & 50 & 50 & 50 & 50 & 50 & 50\\
\hline
\end{tabular}
\label{tableFullyHyperparameters}
\end{table}

\subsubsection{Deep neural networks}

Since neural networks can be trained with additional data without being trained from scratch, they are more versatile compared to other methods such as decision trees. In this paper, a fully-connected neural network architecture (Keras-version 2.2.0 using Tensorflow backend-version 1.8.0) is chosen to approximate the TLE estimation scheme (Figure~\ref{figure9}). The fully-connected neural network learns the mapping between the input and output by connecting all neurons available, and it has 133,801 trainable parameters (total number of weights and biases). It requires a 1D input vector; therefore, 7x50 input matrices are flattened into 1D input vectors with 350 elements (Figure~\ref{figure2}). Unlike decision trees, neural networks are sensitive to the scaling of the input data; therefore, the input data are either standardized (Eq.~\ref{eq:standardizeNN}) or normalized (Eq.~\ref{eq:normalizeNN}) based on the resultant performance of the model. Table~\ref{tableMinMaxScaling} shows the parameters that are used to standardize and normalize the input data. For the  fully-connected machine learning models considered in this work, it is found that the standardization outperforms the normalization.
It should be noted that the parameters in Table~\ref{tableMinMaxScaling} are obtained from only the training data not to create a bias in the machine learning model. The memory restriction is not an issue for training neural networks due to their versatility. The same data excluding the ones with missing values (not a number (NaN) values due to numerical artifacts) that are used to train the gradient boosting trees are used to compare the performances of the two different machine learning methods. XGBoost can be trained with missing values whereas neural networks cannot; therefore, NaN values (0.005\% of the total input data) are removed from the input data. Table~\ref{tableFullyHyperparameters} shows the hyperparameters of the fully-connected neural network that are optimized for neural network models. The Adam optimizer is chosen due to its computational efficiency for problems with large amount of data, and it computes the moving average of the gradients and squared gradients. The parameters of the Adam optimizer that control the decay rate of the moving averages~\cite{KA2015} are $\beta_{1}$ and $\beta_{2}$. The learning rate controls the step size for the Adam optimizer while the decay parameter decays the learning rate at each epoch. The hyperparameters of the fully-connected neural networks are tuned empirically by considering the bias-variance tradeoff to ensure the trained models have generalization capability (Table~\ref{tableFullyBiassVariance}). The grid search methods are not preferred because the models are trained with large amount of data.

\begin{table}[!htbp]
\center
\caption{Bias-Variance tradeoff for fully-connected neural network}
\begin{tabular}{lll}
\hline
Models&Training Squared Mean Error&Test Squared Mean Error\\
\hline
$B^{*}$ & 1.4e-6  & 2.4e-6\\
$i$ & 1.3e-4 & 9.8e-5\\
$\Omega$ & 0.02 & 0.01 \\
$e$ & 1.8e-7 & 2e-7 \\
$\omega$ & 0.24 & 0.22\\
$m$ & 0.2 & 0.3\\
$n$ & 4.6e-9  & 5.3e-9\\
\hline
\end{tabular}
\label{tableFullyBiassVariance}
\end{table}

\begin{equation}
\label{eq:normalizeNN}
z_{n} = \frac{x-min(x)}{max(x)-min(x)}
  \end{equation}
  
\begin{equation}
\label{eq:standardizeNN}
z_{s} = \frac{x-\mu_{(x)}}{\sigma_{(x)}}
  \end{equation}
  
\noindent where $x$ are the input data.

\section{Results and Discussions}

\subsection{Monte-Carlo Approach}

\begin{figure}[!htbp]
\begin{center}
\includegraphics*[width=6cm]{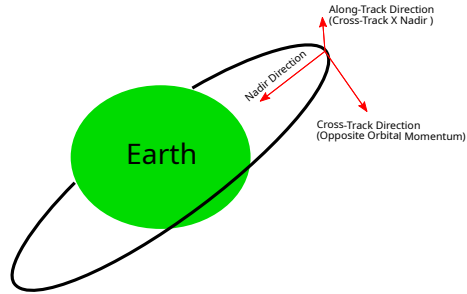}
\end{center}
\caption{The Vehicle Velocity, Local Horizontal (VVLH) reference frame}
\label{figureVVLH}
\end{figure}

A novel approach to estimate a TLE from a given ephemeris without any initial estimate of the TLE using a Monte-Carlo method is introduced. Three particular cases with varying area-to-mass ratios, orbital characteristics and solar activity phases, where the standard differential corrections with the nonlinear least squares method~\cite{VC2008} cannot generate a valid TLE, are chosen as test cases. Initial orbital parameters for the test cases are propagated forward in time using a three-degree-of-freedom (3-DOF) numerical orbit propagator over 1 day. The orbital evolution is defined in the Vehicle Velocity, Local Horizontal (VVLH) reference frame (Figure~\ref{figureVVLH}).The perturbations included are: air drag (Harris-Priester model), the Earth's aspherical gravitational potential with order and degree 20 (Holmes-Featherstone model), solar radiation pressure (Lambertian sphere model), and Sun and Moon third-body gravitational perturbations. Parameters of the test cases used for the numerical propagation are given in Table~\ref{tableOrbitParamMonte}. In addition, the statistical description for the performance of the proposed method in terms of the number of loops and root mean square error in the orbital evolution during both inner and outer loops are provided in Table~\ref{tableStatisticalDescription}. For each test case, 100 TLEs are computed using the proposed method.

\begin{table}[!htbp]
\center
\caption{The parameters of the test cases used for the numerical propagator}
\begin{tabular}{llll}
\hline
Parameters&Test case \#1&Test case \#2&Test case \#3\\
\hline
Epoch (UTC) & 2018-12-12T18:47 & 2014-01-15T10:37 & 2009-08-03T15:18\\
Period (min) & $99.71$ & $94.65$ & $97.32$\\
Mass (kg) & $800$ &$3$ &$1$\\
Area ($m^{2}$) & $6$ &$0.03$ &$0.1$\\
$n$ ($\frac{rad}{s}$) & $0.00105$ &$0.001106$ & $0.001076$\\
$i$ (deg) & $101.45$ &$85.12$ & $97.60$\\
$\Omega$ (deg) & $53.03$ &$90.16$ & $253.29$\\
$\omega$ (deg) & $9.33$ &$204.08$ & $235.32$\\
$m$ (deg) & $359.44$ &$196.9$ & $178.81$\\
$e$ & $0.022353$ &$0.011556$ & $0.025822$\\
$C_{D}$ & $2.2$ &$1.5$ & $0.8$\\
\hline
\end{tabular}
\label{tableOrbitParamMonte}
\end{table}

\begin{table}[!htbp]
\center
\caption{The statistical description for the performance of the proposed method in terms of the number of loops and the root mean square error}
\begin{tabular}{cccccccccc}
\toprule
Parameters &  \multicolumn{2}{c}{Test case \#1} & \multicolumn{2}{c}{Test case \#2}& \multicolumn{2}{c}{Test case \#3}\\
\midrule
{}   & $\mu$   & $\sigma$    & $\mu$   & $\sigma$ & $\mu$   & $\sigma$\\
Final RMSE (km)   &  4.92 & 2.92   & 4.95  & 2.76 & 4.98  & 2.82\\
Inner loop number   &  54.39 & 50.22   & 43.91  & 42.38 & 55.28  & 58.06\\
Inner loop RMSE (km)   &  659.49  &  482.67   & 556.65  & 416.40 & 619.22  & 447.76\\
Outer loop number   &  1.43  &  0.83   & 1.41  & 0.71 & 1.38  & 0.74\\
Outer loop RMSE (km)  &  7.0  &  4.33   & 7.26  & 4.4 & 7.0  & 4.17\\
\hline
\end{tabular}
\label{tableStatisticalDescription}
\end{table}

\begin{figure}[!htbp]
\begin{center}
\includegraphics*[width=10cm]{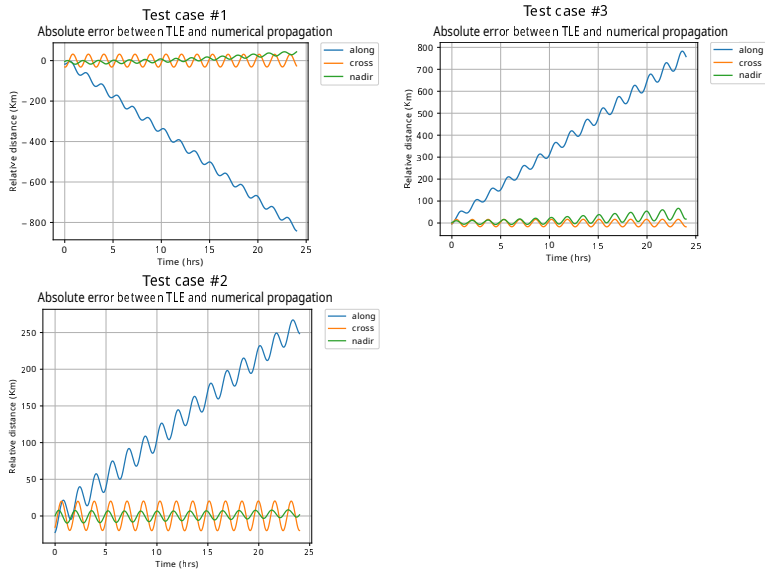}
\end{center}
\caption{Absolute relative distance with respect to numerically propagated orbit when Keplerian elements are used as TLE}
\label{figure6}
\end{figure}

\begin{figure}[!htbp]
\begin{center}
\includegraphics*[width=10cm]{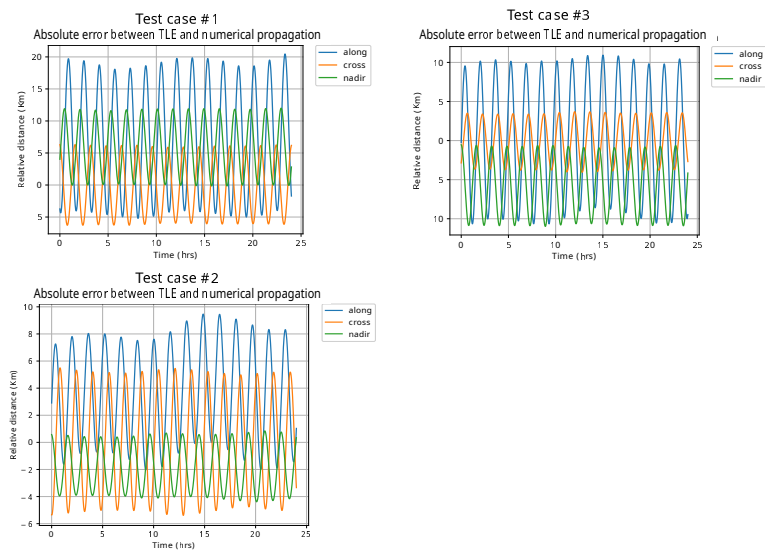}
\end{center}
\caption{Absolute relative distance between the numerically propagated orbit and the SGP4 propagated orbit when the inner loop is satisfied}
\label{figure7}
\end{figure}

\begin{figure}[!htbp]
\begin{center}
\includegraphics*[width=10cm]{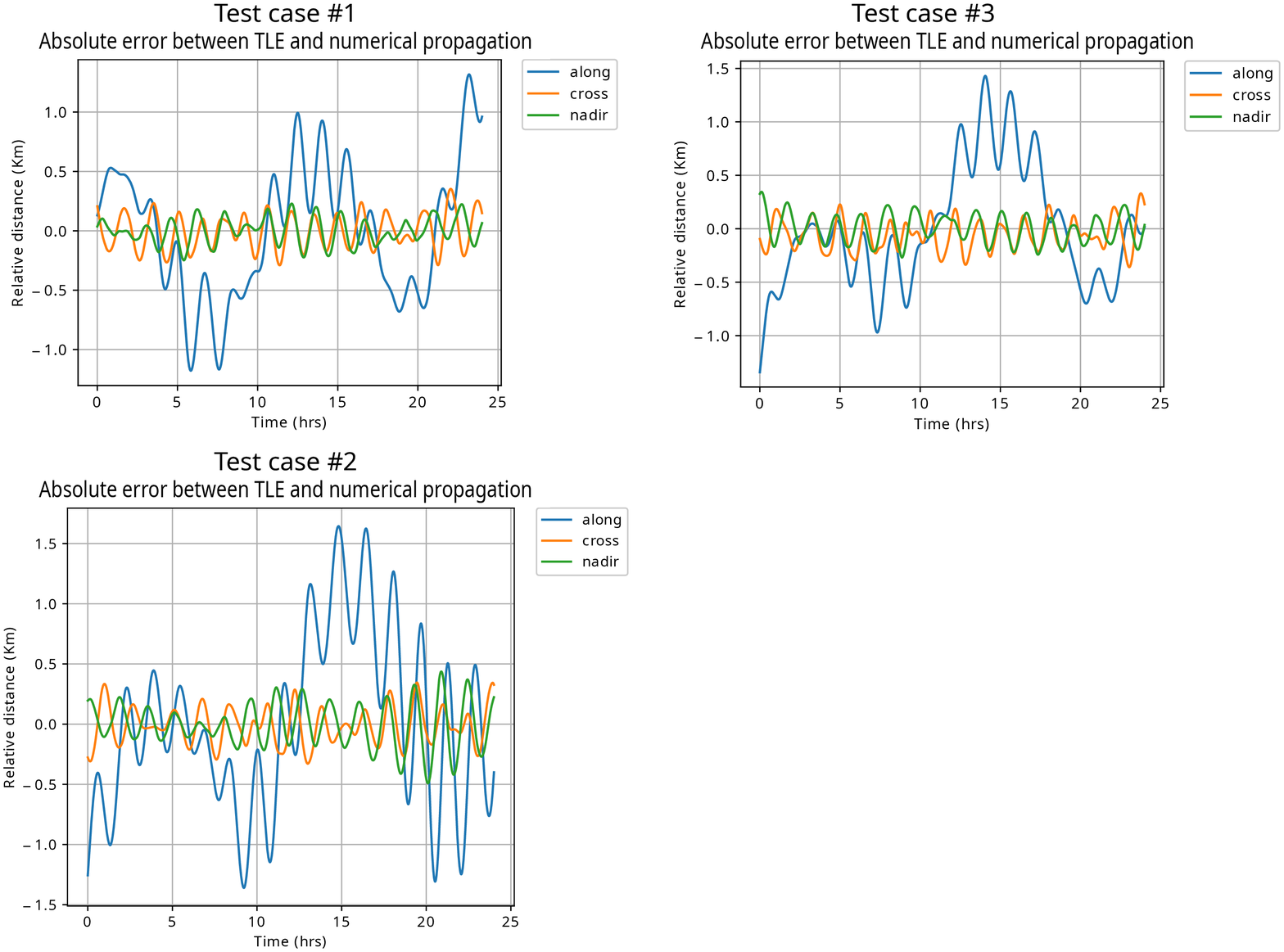}
\end{center}
\caption{Absolute relative distance between the numerically propagated orbit and the SGP4 propagated orbit when a valid TLE is generated}
\label{figure8}
\end{figure}

Figure~\ref{figure6} shows the absolute errors in the orbital evolution between the orbit propagated by SGP4 using the Keplerian elements and the TLE obtained from the Combined Space Operations Center (CSpOC). The most significant error is in the along-track direction while the second largest error is in the cross-track direction for all test cases. The drastic divergence of orbital evolution leads to large corrections to the initial TLE estimate (Keplerian elements), and this results in divergence in the standard TLE estimation scheme~\cite{VC2008}. The nonlinear least squares and differential corrections method minimize the mean square errors. The secular error growth results in TLEs with large errors. Therefore, an efficient algorithm that searches for bounded error evolution is required. Because a valid TLE should fit the whole orbit, an osculating TLE at the epoch time that fits the whole orbit is searched for using the Monte-Carlo method. The inner loop stopping criteria is selected as  RMSE of 30 km (Figure~\ref{figure1}). The outer loop is terminated when the standard TLE estimation scheme~\cite{VC2008} computes a valid TLE (RMSE of 10 km and smaller for this work). Figure~\ref{figure7} shows the absolute relative distance over one day when the inner loop criterion is satisfied. Figure~\ref{figure8} shows the absolute relative distance when a valid TLE is computed. The above TLE satisfies both the inner and the outer loops simultaneously (Figure~\ref{figure1}).

In conclusion, the present Monte-Carlo method can estimate a TLE without any initial estimate. Such capability is beneficial to space situational awareness (SSA) because TLEs are required for establishing first contact with the spacecraft during the Launch and Early Orbit Phase (LEOP). Most ground stations have propriety software that requires TLEs to track the satellites. Although the proposed method addresses the difficulty related to unavailability of the initial estimate, the inner loop is iterated 50 times on the average to find an initial estimate with desired precision. Therefore, the feasibility of using machine learning methods to generate initial estimates with the desired precision is investigated in the following sections. 

\subsection{Machine Learning Models}

In this section, the performances of the gradient boosting trees and the fully-connected neural networks that are trained to map an orbital evolution to the associated TLE are presented using two different approaches. For the first performance test, the metric is chosen as the absolute difference between the TLEs that are predicted by machine learning models and the CSpOC TLEs. The first test is also used to determine the best combinations of machine learning models which is selected for the second test. For the second performance test, the metric is chosen as the absolute relative distance in the VVLH reference frame  between the positions propagated by the estimated TLEs and the CSpOC TLEs.

\subsubsection{$B^{*}$ machine learning model}

\begin{figure}[!htbp]
\begin{center}
\includegraphics*[width=8cm]{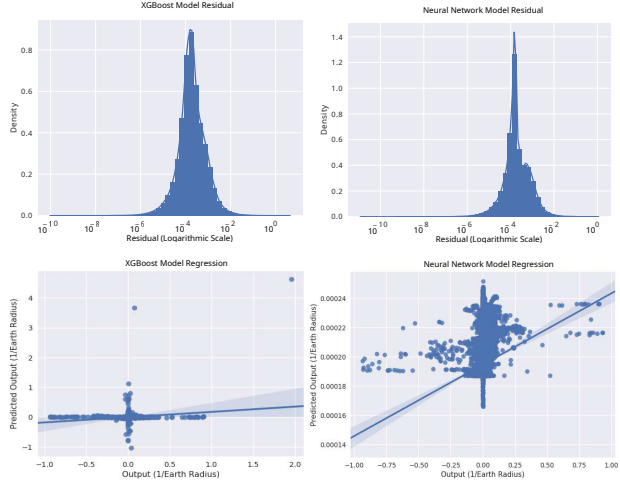}
\end{center}
\caption{Performance of $B^{*}$ machine learning models}
\label{figureBstarModelsPerformance}
\end{figure}

The absolute residual plot and the scatter plot with regression of the predictions of the $B^{*}$ using machine learning models are presented in Figure~\ref{figureBstarModelsPerformance}. Both machine learning models have difficulties in determining the decision boundary that can accurately map the features to the target values, and this may be related to the modification of $B^{*}$ to account for perturbations~\cite{DV2013}. There are  outliers observed in the $B^{*}$ data that are kept  (99.9\% of $B^{*}$ values in the test data are between -0.05 and 0.05 (1/Earth radius)) which may affect the machine learning models. The poor performance of the models can be attributed to the modification of the $B^{*}$ value during the estimation process as the tree-based models are insensitive to the outliers in the data. Both methods output values close to the average of all target values ($10^{-4}$). The 48\% of all $B^{*}$ values in the test data are between 0.0001 and 0.0009 (1/Earth radius).  The percentage of residuals smaller than 0.0001 is 28.2\% for the gradient boosting model and 25.1\% for neural network model. The regression plots (Figure~\ref{figureBstarModelsPerformance}) show that the performance of both machine learning models is similar for different values of $B^{*}$. Therefore, the gradient boosting tree $B^{*}$ model is chosen as the best model irrespective of the $B^{*}$ values because the percentage of residuals smaller than 0.0001 is higher for the gradient boosting trees.

\subsubsection{Mean motion machine learning model}

\begin{figure}[!htbp]
\begin{center}
\includegraphics*[width=8cm]{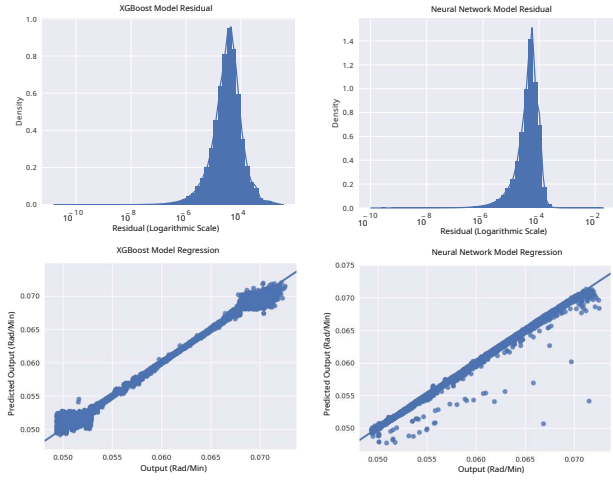}
\end{center}
\caption{Performance of mean motion ($n$) machine learning models}
\label{figureNoModelsPerformance}
\end{figure}

The absolute residual plot and the scatter plot with regression of the predictions of mean motion ($n$) using machine learning models are presented in Figure~\ref{figureNoModelsPerformance}. The residuals smaller than 0.0001 (rad/min) are 86.2\% of the test data for the gradient boosting tree model and 89.4\% for fully-connected network model. The above criterion of 0.0001 (rad/min) is the RMSE of the predictions of the test data. The regression plot of the gradient boosting tree model in Figure~\ref{figureNoModelsPerformance} shows that the variance becomes larger for values smaller than 0.053 (rad/min) and bigger than 0.067 (rad/min), and the regression plot for the fully-connected neural network shows that there are outliers in the predictions. Table~\ref{tableMeanMotionClose} presents the mean and variance of residuals for different mean motion values. The fully-connected neural network outperforms the gradient boosting trees regarding prediction accuracy for all values of mean motion. Therefore, the fully-connected neural network model is chosen as the best machine learning model for predicting mean motion. For the orbital evolution accuracy test at the end of the section, the fully-connected neural network model is used to predict mean motion unless the absolute difference between the prediction of neural network and the prediction of the gradient boosting trees are larger than 0.0002 (rad/min) for values of $n$ between 0.053 (rad/min) and 0.067 (rad/min) to account for outliers present in the predictions of the fully-connected neural network model (Figure~\ref{figureNoModelsPerformance}). The above threshold value of 0.0002 (rad/min) is $\mu+3\sigma$ of the residuals for the fully-connected neural network for values of $n$ between 0.053 (rad/min) and 0.067 (rad/min).

\begin{table}[!htbp]
\center
\caption{Mean and variance of residuals for different $n$ values}
\begin{tabular}{llll}
\hline
Mean and Variance & $n<0.053$ & $n>0.067$ & $0.053<n<0.067$\\
\hline
Mean (XGBoost) & 0.0006 & 0.0003 & 6.5e-5\\
Mean (Neural Network) & 7.2e-5 & 7e-5 & 5.2e-5\\
Variance (XGBoost) & 3.2e-7 & 9.4e-8 & 1.8e-8\\
Variance (Neural Network) & 8.5e-9 & 2.5e-8 & 2.6e-9\\ 
\hline
\end{tabular}
\label{tableMeanMotionClose}
\end{table}

\subsubsection{Inclination machine learning model}

\begin{figure}[!htbp]
\begin{center}
\includegraphics*[width=8cm]{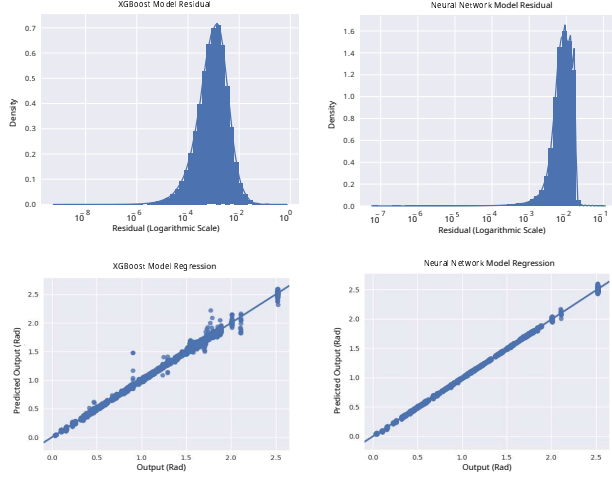}
\end{center}
\caption{Performance of inclination ($i$) machine learning models}
\label{figureIncloModelsPerformance}
\end{figure}

The absolute residual plot and the scatter plot with regression of the predictions of inclination ($i$) using machine learning models are presented in Figure~\ref{figureIncloModelsPerformance}. For the gradient boosting trees, the mean of the residuals is 0.002 radians, and the variance of the residuals is 1.6e-5 square radians. For the fully-connected neural network, the mean of the residuals is 0.008 radians, and the variance of the residuals is 1.9e-5 square radians. The gradient boosting tree model is chosen as the best machine learning model for predicting inclination. However, the regression plot for the gradient boosting trees shows that there are outliers in the predictions. Therefore, the gradient boosting tree model is used to predict inclination unless the absolute difference between the prediction of neural network and the prediction of the gradient boosting trees are larger than 0.014 (rad), which is $\mu+3\sigma$ of the residuals for the gradient boosting trees, for the orbital evolution accuracy test at the end of the section.

\subsubsection{Right ascension of the ascending node machine learning model}

\begin{figure}[!htbp]
\begin{center}
\includegraphics*[width=8cm]{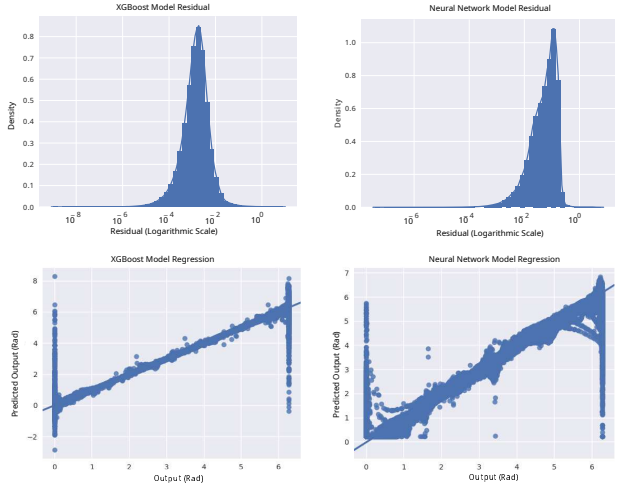}
\end{center}
\caption{Performance of right ascension of the ascending node ($\Omega$) machine learning models}
\label{figureRaanModelsPerformance}
\end{figure}

The absolute residual plot and the scatter plot with regression of the predictions of right ascension of the ascending node ($\Omega$) using machine learning models are presented in Figure~\ref{figureRaanModelsPerformance}. For the gradient boosting trees, the mean of the residuals is 0.0045 radians, and the variance of the residuals is 0.0029 square radians. For the fully-connected neural network, the mean of the residuals is 0.076 radians, and the variance of the residuals is 0.0115 square radians. The equinoctial element $Lv$ ($\Omega+\omega+\nu$) is in radians in the features. This  leads to the large variances in the residuals of the predictions of both methods because the machine learning models fail to recognize the cyclic behavior when any of the mean elements, namely $\Omega$, $\omega$, $\nu$, switch values between $2\pi$ and $0$ as the orbits evolve in time. The advancement in the values of the above mentioned mean elements between $2\pi$ and $0$ occurs for the $0.08\%$ of the test data. This issue will be addressed in future studies by introducing additional features into the features, such as $\Omega$, $\omega$, $\nu$, and applying data transformations. The gradient boosting tree model is chosen as the best machine learning model for predicting $\Omega$ because the mean of the residuals for its predictions is one order magnitude lower than that of the fully-connected neural network.

\subsubsection{Argument of the perigee and mean anomaly machine learning models}

\begin{figure}[!htbp]
\begin{center}
\includegraphics*[width=8cm]{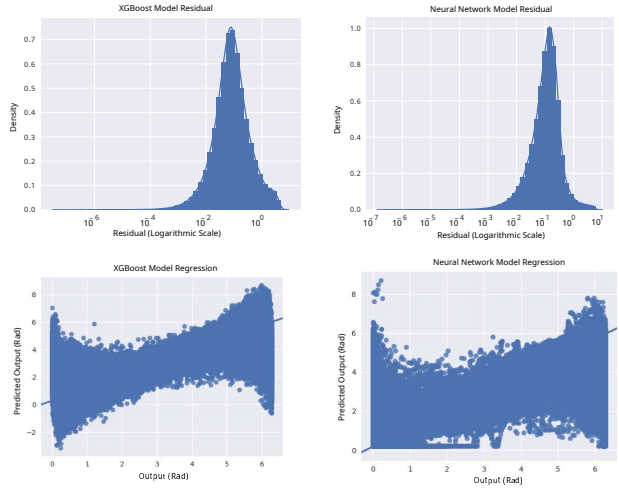}
\end{center}
\caption{Performance of argument of perigee ($\omega$) machine learning models}
\label{figureArgpoModelsPerformance}
\end{figure}

\begin{figure}[!htbp]
\begin{center}
\includegraphics*[width=8cm]{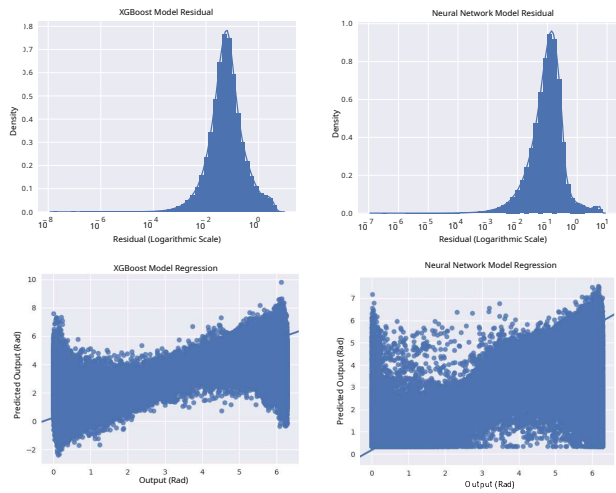}
\end{center}
\caption{Performance of mean anomaly ($m$) machine learning models}
\label{figureMoModelsPerformance}
\end{figure}

\begin{figure}[!htbp]
\begin{center}
\includegraphics*[width=8cm]{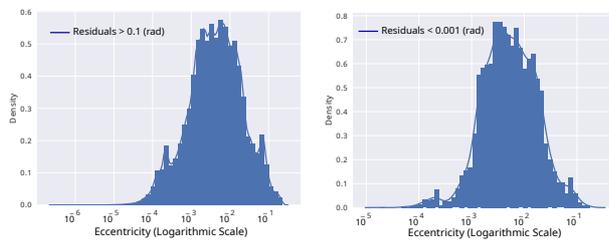}
\end{center}
\caption{Distribution of residuals with respect to eccentricities}
\label{figureEccoVersusResidual}
\end{figure}

The absolute residual plots and the scatter plots with regression of the predictions of the argument of perigee ($\omega$) and the mean anomaly ($m$) using machine learning models are presented in Figure~\ref{figureArgpoModelsPerformance} and Figure~\ref{figureMoModelsPerformance} respectively. Machine learning models for both mean elements are analyzed together due to their similar performances in the residuals of their predictions. Table~\ref{tableMeanAnomalyArgofperigee} shows the mean and variance of the residuals of the predictions for the machine learning models. According to the Table~\ref{tableMeanAnomalyArgofperigee}, the gradient boosting tree model is chosen as the best machine learning model for predicting argument of perigee ($\omega$), and the fully-connected neural network is chosen as the best machine learning model for predicting the mean anomaly ($m$). However, both the argument of perigee and the mean anomaly machine learning models have relatively poor performance in determining the decision boundary that can map the orbital evolution to the mean elements accurately. In addition to the difficulty of capturing the cyclic behavior of angles close to $0$ and $2\pi$ due to $Lv$ feature (as discussed for $\Omega$ above), all machine learning models learn a representation that has large variations in the residuals of its predictions, and this may indicate that the mean anomaly and the argument of perigee are modified to fit the perturbed orbit as it is the case for $B^{*}$. In addition, there is no significant relationship between the eccentricity and the accuracy of the predictions of the argument of perigee (Figure~\ref{figureEccoVersusResidual}). In Figure~\ref{figureEccoVersusResidual}, the distributions of the eccentricities for the residuals of predictions of $\omega$ for values smaller than $0.001$ and larger than $0.1$ are presented. The main reason for the poor performance of the models for the parameters is the fact that the sum of these two parameters define the location of the object in the orbit. Therefore, the estimator (non-linear least squares with differential corrections method) changes each parameter arbitrarily to reduce the sums of the squared errors without changing the actual location of the object in the orbit because most LEO objects have small eccentricities, and this behavior is observed for all test cases in Section 3.1. 

\begin{table}[!htbp]
\center
\caption{Mean and variance of residuals for the mean anomaly and the argument of perigee}
\begin{tabular}{lll}
\hline
Machine learning models & Mean ($rad$) & Variance ($rad^{2}$)\\
\hline
Mean Anomaly (XGBoost) & 0.19 & 0.23 \\
Mean Anomaly (Neural Network) & 0.21 & 0.26 \\
Argument of Perigee (XGBoost) & 0.24 & 0.27 \\
Argument of Perigee (Neural Network) & 0.20 & 0.18 \\ 
\hline
\end{tabular}
\label{tableMeanAnomalyArgofperigee}
\end{table}

\subsubsection{Eccentricity machine learning model}

\begin{figure}[!htbp]
\begin{center}
\includegraphics*[width=8cm]{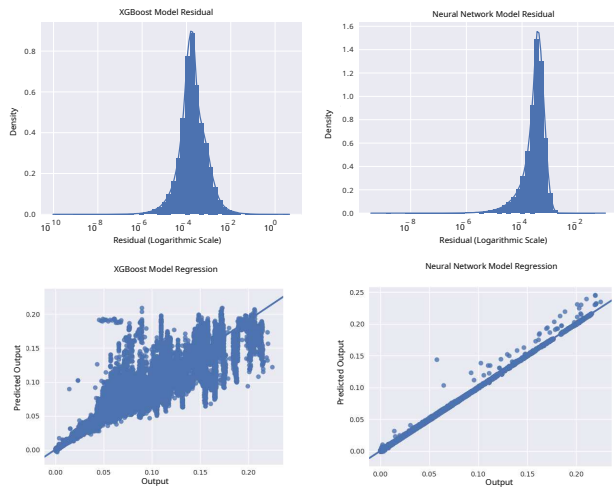}
\end{center}
\caption{Performance of eccentricity ($e$) machine learning models}
\label{figureEccoModelsPerformance}
\end{figure}

The absolute residual plot and the scatter plot with regression of the predictions of eccentricity ($e$) using machine learning models are presented in Figure~\ref{figureEccoModelsPerformance}. For the gradient boosting trees, the mean of the residuals is 0.0008, and the variance of the residuals is 1.23e-5. For the fully-connected neural network, the mean of the residuals is 0.0003, and the variance of the residuals is 7.59e-8. The fully-connected neural network is chosen as the best machine learning model.

\subsubsection{The performance evaluation of the selected best machine learning models}

The performances of machine learning methods differ based on the features selected. Two different machine learning models are trained to predict each mean elements of a TLE, and the best performing model is chosen to be validated by the errors in the orbital evolution. The error in the orbital evolution is computed by propagating TLEs estimated by the selected best machine learning models and the associated CSpOC TLEs backward in time using SGP4 for 50 orbital periods (Figure~\ref{figure2}). Table~\ref{tableBestModels} summarizes the best performing models for each mean element that are discussed in detail above. 

\begin{table}[!htbp]
\center
\caption{Best machine learning models for predicting mean elements of a TLE}
\begin{tabular}{|p{3cm}|p{9cm}|}
\hline
Mean elements&Best machine learning models\\
\hline
\hline
$B^{*}$&Gradient boosting trees\\
\hline
$n$&Fully-connected neural network (unless the absolute residual between two models are larger than 0.0002 rad/min for values of $n$ between 0.053 rad/min and 0.067 rad/min)\\
\hline
$i$&Gradient boosting trees (unless the absolute residual between two models are larger than 0.014 rad)\\
\hline
$\Omega$&Gradient boosting trees\\
\hline
$\omega$&Gradient boosting trees\\
\hline
$m$&Fully-connected neural network\\
\hline
$e$&Fully-connected neural network\\
\hline
\end{tabular}
\label{tableBestModels}
\end{table}

\begin{figure}[!htbp]
\begin{center}
\includegraphics*[width=10cm]{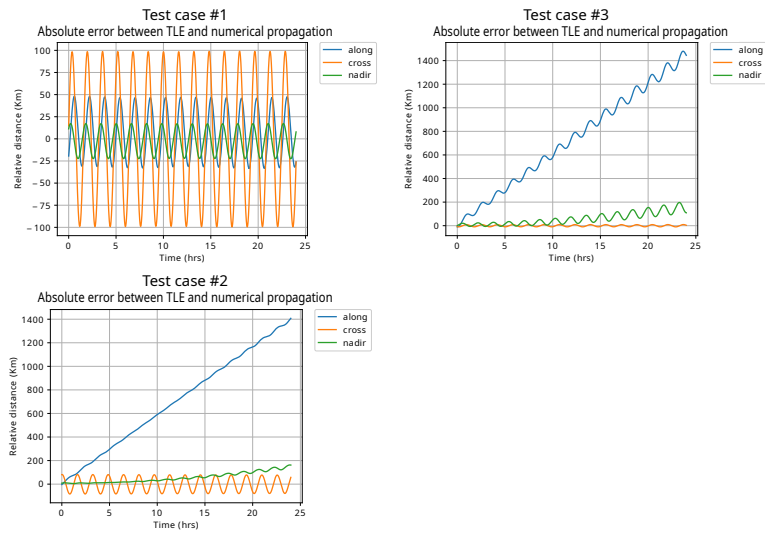}
\end{center}
\caption{Absolute relative distances with respect to numerically propagated orbits when TLEs estimated by machine learning models are used}
\label{figureMLfirst}
\end{figure}

\begin{figure}[!htbp]
\begin{center}
\includegraphics*[width=10cm]{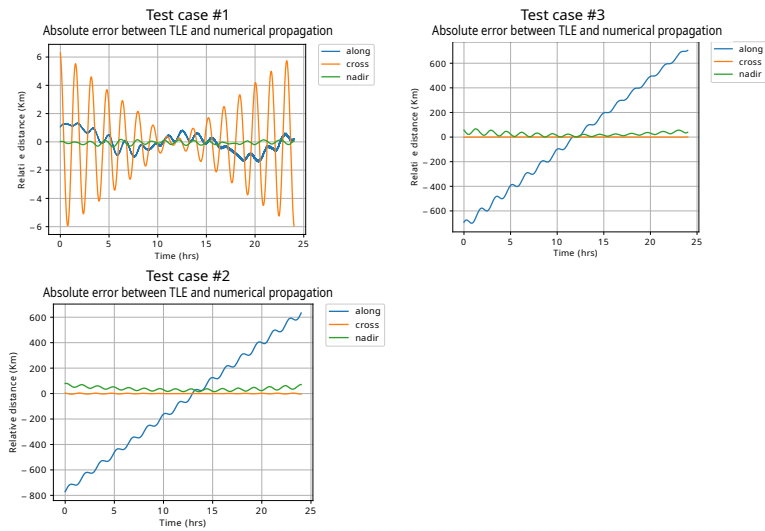}
\end{center}
\caption{Absolute relative distances between the numerically propagated orbits and the SGP4 propagated orbits when the initial estimates that are computed by the machine learning models are improved by nonlinear least squares with differential corrections method}
\label{figureMLfinal}
\end{figure}

The performance of the machine learning models is evaluated by replacing the inner loop of the Monte-Carlo approach, which searches for an initial estimate with desired precision, with the selected best machine learning models. For all three test cases (Table~\ref{tableOrbitParamMonte}), the orbital evolution (Figure~\ref{figure2}) associated with the desired TLE is feed into the selected best machine learning models to obtain an initial estimate. The osculating Keplerian mean anomaly and argument of perigee are replaced with the counterparts estimated by the machine learning models because of the poor performance of the models for these two parameters.The selected best machine learning models are able to provide an initial estimate with desired precision only for the test case \#1, which is a low area-to-mass ratio space object. For the other test cases, the estimated TLEs have RMSE values over 100 km, and this indicates that the precision that is required for mapping the orbital evolution to the associated initial estimate of the desired TLE should be higher. Figure~\ref{figureMLfirst} shows the absolute errors in the orbital evolution between the orbit propagated using SGP4 with TLEs estimated by the selected best machine learning models as compared to the TLE obtained from the CSpOC. Figure~\ref{figureMLfinal} shows the absolute relative errors between the numerically propagated orbits and the orbits propagated by SGP4 when the initial estimates that are computed by the machine learning models are improved by nonlinear least squares and differential corrections methods.

In conclusion, the present best machine learning models (selected based on their performance) that are trained with TLEs obtained from the official US space catalog have the potential to provide an initial estimate to estimate a TLE with desired precision. The inner loop (Figure~\ref{figure2}), which searches for the global minimum by using brute force approach, in the Monte-Carlo TLE estimation method can be replaced by machine learning models because there is no secular error growth in the orbital evolution of TLEs predicted by machine learning models (test case \#1). Therefore, the standard differential corrections with nonlinear least squares method~\cite{VC2008} can converge to a local minimum, and the computation time can be reduced significantly.

\section{Conclusions and Future Work}

In this paper, a new Monte-Carlo TLE estimation method that does not require an initial estimate of the desired TLE is developed, and the feasibility of approximating the inverse mapping of publicly available SGP4 algorithm~\cite{VC2006} for LEO objects using publicly available TLEs and state-of-the-art machine learning methods is investigated. The present Monte-Carlo TLE estimation method can estimate a TLE without any initial estimate from an orbital evolution with an RMSE value smaller than 1 km over 1 day for space objects with varying area-to-mass ratios and orbital characteristics. The selected best machine learning models are shown to provide an initial estimate with desired precision for the associated TLE of a low area-to-mass ratio object.


To the author's knowledge, this work is the first effort to approximate a mapping between orbital evolution and the TLE parameters using machine learning. There are a couple of benefits of developing such mappings for SSA. First, the computational power that is required to estimate TLEs on a daily basis can be used for other important tasks. Second, the capability of reducing the dimension of orbital evolution down to mean elements using machine learning models provides valuable insights in utilizing them for orbit determination. The initial results of machine learning models that can predict TLEs are promising. The TLEs predicted by the machine learning model can determine the reference orbit with the desired precision. Additional input parameters, non-discontinuous transformations of cyclic parameters, and different machine learning architectures could enhance the capabilities of models. Developing different machine learning models for different orbital regimes in LEO can improve the current prediction accuracy of the models as well. For future studies, the feasibility of enhancing the predictions of the right ascension of the ascending node and the inclination, and the location of the space object within the reference trajectory using physics-embedded machine learning models will be investigated.










\section*{Conflict of interests}

On behalf of all authors, the corresponding author states that there is no conflict of interest. 

\bibliographystyle{spphys}       

\bibliography{sample}

\begin{thebibliography}{10}
\providecommand{\url}[1]{{#1}}
\providecommand{\urlprefix}{URL }
\expandafter\ifx\csname urlstyle\endcsname\relax
  \providecommand{\doi}[1]{DOI \discretionary{}{}{}#1}\else
  \providecommand{\doi}{DOI \discretionary{}{}{}\begingroup
  \urlstyle{rm}\Url}\fi

\bibitem{DV2013}
D.A. Vallado, \emph{Fundamentals of astrodynamics and applications}, 4th edn.
  (Microcosm Press, 2013)

\bibitem{VC2006}
D.~Vallado, P.~Crawford, R.~Hujsak, T.~Kelso, Revisiting spacetrack report\# 3
  p. 6753 (2006)

\bibitem{VC2008}
D.~Vallado, P.~Crawford, Sgp4 orbit determination p. 6770 (2008)

\bibitem{JGMK1996}
E.~Jochim, E.~Gill, O.~Montenbruck, M.~Kirschner, Gps based onboard and
  onground orbit operations for small satellites, Acta Astronautica
  \textbf{39}(9-12), 917 (1996)

\bibitem{LBS2002}
B.~Lee, Norad tle conversion from osculating orbital element, Journal of
  Astronomy and Space Sciences \textbf{19}(4), 395 (2002)

\bibitem{MG2000}
O.~Montenbruck, E.~Gill, Real-time estimation of sgp4 orbital elements from gps
  navigation data pp. 26--30 (2000)

\bibitem{GOH2018}
S.T. Goh, K.S. Low, Real-time estimation of satellite's two-line elements via
  positioning data pp. 1--7 (2018)

\bibitem{BLD2015}
H.~Bolandi, M.~Ashtari~Larki, S.~Sedighy, M.~Zeighami, M.~Esmailzadeh,
  Estimation of simplified general perturbations model 4 orbital elements from
  global positioning system data by invasive weed optimization algorithm,
  Proceedings of the Institution of Mechanical Engineers, Part G: Journal of
  Aerospace Engineering \textbf{229}(8), 1384 (2015)

\bibitem{FS1995}
Y.~Freund, R.E. Schapire, A decision-theoretic generalization of on-line
  learning and an application to boosting, Journal of computer and system
  sciences \textbf{55}(1), 119 (1997)

\bibitem{KJ2013}
M.~Kuhn, K.~Johnson, \emph{Applied predictive modeling}, vol.~26 (Springer,
  2013)

\bibitem{Fr2001}
J.H. Friedman, Greedy function approximation: a gradient boosting machine,
  Annals of statistics pp. 1189--1232 (2001)

\bibitem{Fr2002}
J.H. Friedman, Stochastic gradient boosting, Computational statistics \& data
  analysis \textbf{38}(4), 367 (2002)

\bibitem{CG2016}
T.~Chen, C.~Guestrin, Xgboost: A scalable tree boosting system pp. 785--794
  (2016)

\bibitem{Cy1989}
G.~Cybenko, Approximation by superpositions of a sigmoidal function,
  Mathematics of control, signals and systems \textbf{2}(4), 303 (1989)

\bibitem{HSW1989}
K.~Hornik, M.~Stinchcombe, H.~White, Multilayer feedforward networks are
  universal approximators, Neural networks \textbf{2}(5), 359 (1989)

\bibitem{GYC2016}
I.~Goodfellow, Y.~Bengio, A.~Courville, \emph{Deep learning} (MIT press, 2016)

\bibitem{MVP2010}
L.~Maisonobe, V.~Pommier, P.~Parraud, Orekit: an open-source library for
  operational flight dynamics applications pp. 3--6 (2010)

\bibitem{KA2015}
D.P. Kingma, J.~Ba, Adam: A method for stochastic optimization  (2015)

\end{thebibliography}

\end{document}